\documentclass{nature}

\usepackage[utf8]{inputenc}
\usepackage{xcolor}
\usepackage{amsmath}
\newcommand{\Rev}[1]{\textcolor{black}{ #1}}

\usepackage{braket}
\usepackage{enumitem}
\usepackage[mathscr]{euscript}
\usepackage{graphicx}
\usepackage{caption}
\usepackage{enumitem}
\usepackage{lineno}
\usepackage{comment}
\makeatletter
\let\saved@includegraphics\includegraphics
\AtBeginDocument{\let\includegraphics\saved@includegraphics}
\renewenvironment*{figure}{\@float{figure}}{\end@float}
\makeatother

\title{Rotatum of Light}

\author{Ahmed H. Dorrah$^{1,*}$, Alfonso Palmieri$^1$, Lisa Li$^1$, and Federico Capasso$^{1,*}$} 

\begin{document}
\captionsetup[figure]{labelfont={bf},name={Figure},labelsep=colon,font=small}

\maketitle

\begin{affiliations}
 \item Harvard John A. Paulson School of Engineering and Applied Sciences, Harvard University, Cambridge, Massachusetts 02138, USA \\
  
{*}Corresponding author(s): dorrah@seas.harvard.edu;\ capasso@seas.harvard.edu
\end{affiliations}

\begin{abstract}
Vortices are ubiquitous in nature and can be observed in fluids, condensed matter, and even in the formation of galaxies. Light, too, can evolve like a vortex. Optical vortices are exploited in light-matter interaction, free-space communications, and imaging. Here, we introduce optical rotatum; a new degree-of-freedom of light in which an optical vortex experiences a quadratic chirp in its orbital angular momentum along the optical path. We show that such an adiabatic deformation of topology is associated with the accumulation of a Berry phase factor which in turn perturbs the propagation constant (spatial frequency) of the beam. Remarkably, the spatial structure of optical rotatum follows a logarithmic spiral---a signature that is commonly seen in the pattern formation of seashells and galaxies. Our work expands previous literature on structured light, offers new modalities for light-matter interaction, communications, and sensing, and hints to analogous effects in condensed matter physics and Bose-Einstein condensates.   
\end{abstract}

\section{Introduction}
Vortex flow is a signature of many systems in nature and is often seen in turbulent fluids, smoke rings, tornados, electric and magnetic currents, and even the formation of galaxies~\cite{10.1119/1.14177}. Electromagnetic radiation, including light, can also evolve like a vortex both in space~\cite{Yao11,Rubinsztein-Dunlop_2017} and time~\cite{doi:10.1126/science.aaw9486,Zhang760,Huang1018,Shen_2023}. Optical vortices are typically characterized by an azimuthal phase-dependence of $e^{i\ell\phi}$, where $\ell$ denotes the slope of the phase~\cite{PhysRevA.45.8185}. Such profile carries an on-axis phase singularity which forces the Poynting vector to skew off-axis. This non-zero transverse component of the Poynting vector in turn creates orbital angular momentum (OAM) of $\ell\hbar$ per photon~\cite{OAM1995,OAM1999}. Besides their rich physics, optical vortices have enabled new degrees-of-freedom for light-matter interaction~\cite{Friese1998,Grier2003,Padgett2011} and have been utilized in free space communications~\cite{Wang2012,Willner:15}, remote sensing~\cite{PhysRevLett.110.043601,Cvijetic2015,Dorrah2018}, imaging~\cite{Hell:94,Vicidomini2018}, quantum information processing~\cite{Molina-Terriza2007,Erhard2018,Fang2020}, among many other applications~\cite{Padgett:17,Shen2019}. A variety of tools have been used to generate optical vortices, including digital holography~\cite{Forbes:16,10.1117/1.OE.59.4.041202}, metasurfaces~\cite{Yu333,Devlin896,doi:10.1126/science.abi6860}, spiral~\cite{Beijersbergen1994}, and geometric phase plates~\cite{PhysRevLett.96.163905}. 

OAM is regarded as a conserved quantity under free space propagation and is associated with a quantized topological charge $\ell$. Thus, OAM cannot be freely modified~\cite{jackson_classical_1999,Allen2000,Gbur:08}. Nevertheless, several complex patterns of optical vortices have been reported thanks to an abundance of advanced wavefront shaping tools. For instance, the vorticity of light can now be locally modulated along the optical path~\cite{PhysRevA.93.063864,Davis:s,Yang2018,Zhang2019,Wang2019,Luo:s,Dorrah2021NC}. In this case, it is understood that the OAM density may vary locally (at the center of the beam) while keeping the global OAM conserved at each plane along the propagation direction~\cite{PhysRevA.98.043846,Dorrah2021NC}. Vortices of this kind have been utilized in refractometry by mapping light's rotation to the unknown refractive index~\cite{Dorrah2018}, high-capacity free space communications by transmitting different symbols to different receivers located along the optical path~\cite{8123613}, and in robust information transfer by matching the structured beam to the spatially-varying turbulence profile of the medium~\cite{Zhou2023}.    

Different techniques have been used to spatially modulate the topological charge of optical vortices with propagation. A common strategy relies on interfering multiple co-propagating OAM modes with different $\ell$-values and propagation constants such that their spatial beating produces an envelope that changes its OAM, locally, with propagation~\cite{PhysRevA.93.063864,Davis:s,Dorrah2021NC}. This approach has demonstrated vortex beams that change their topological charge from one integer value to another following a step-like transition. \Rev{In this case, the vortex beam carries integer values of $\ell$ or superpositions thereof}. Continuous evolution of OAM, spanning fractional~\cite{ZhangZeng} and integer $\ell$-values, has also been demonstrated by transmitting light through spiral slits~\cite{Yang2018,Zhang2019,PhysRevApplied.12.064007}. In this case, the linear growth or decay of OAM in space can be controlled by engineering the geometry of the spiral. The temporal analogue of this behavior is referred-to as self-torque of light \Rev{($\mathbf{\tau}$)}, where light's vorticity changes linearly as a function of time giving rise to a non-zero first-order derivative of OAM \Rev{($\textbf{L}_z$), such that $\mathbf{\tau}=d \mathbf{L}_z/d t$} ~\cite{doi:10.1126/science.aaw9486,Shen_2023}. Light of this kind provides an extraordinary tool for laser-matter manipulation on attosecond time and nanometer spatial scales. It also posed a new question: can light change its \Rev{self-}torque with propagation \Rev{(i.e., $d^2 \mathbf{L}_z/d t^2 \neq 0$)} ? While higher order derivatives (second, third...etc) of OAM have been observed and studied in classical mechanics~\cite{goldstein2002classical}, referred to as jerk or rotatum, their electromagnetic analogue has not been introduced in the literature to date despite their rich physical dynamics \Rev{and} potential applications.    

In this work, we reveal a new degree-of-freedom of light which we dub optical rotatum. Optical rotatum describes vortices whose $\ell$-value experiences a quadratic chirp along the optical path; giving rise to a non-zero second-order derivative of OAM \Rev{(i.e., $d^2 \mathbf{L}_z/d z^2 \neq 0$)}---a quantity that has not been observed in electrodynamic systems to date. The mechanism relies on introducing an azimuthally varying gradient in the spatial frequency ($k$-vector) of the beam. \Rev{Hence, in analogy to vortex beams which carry an azimuthal phase gradient, here each point on the azimuth is characterized by a slightly different $k$-vector}. Upon propagation, different points along the azimuthal direction of the wavefront will accumulate different phase delays causing the phasefront to acquire a singularity on axis while experiencing a continuous deformation in its helical twist. By judiciously designing this azimuthal $k$-gradient \Rev{($d k_z/d\phi$)}, it is possible to generate optical vortices whose OAM can locally follow any polynomial dependence (i.e., linear, quadratic, or cubic...etc) along the optical path. Notably, such an adiabtic evolution in OAM gives rise to a topological phase factor (i.e., a Berry phase) which perturbs the propagation constant of the beam and that can be tailored by design. Besides their rich dynamics, light beams of this kind can be used as optical rulers for precise depth sensing and metrology and can also find application in the efficient sorting of colloids in 3D, to name but a few. Our work expands on current literature of structured light generation, hints to similar observations in many other physical systems in nature, and can be applied beyond optics; for e.g., in ultrasonic~\cite{Fueaba9876} and electron beams~\cite{Verbeeck2010}. 

\section{Concept}

\begin{figure}[h!]
    \centering
    \includegraphics[width=0.975\textwidth]{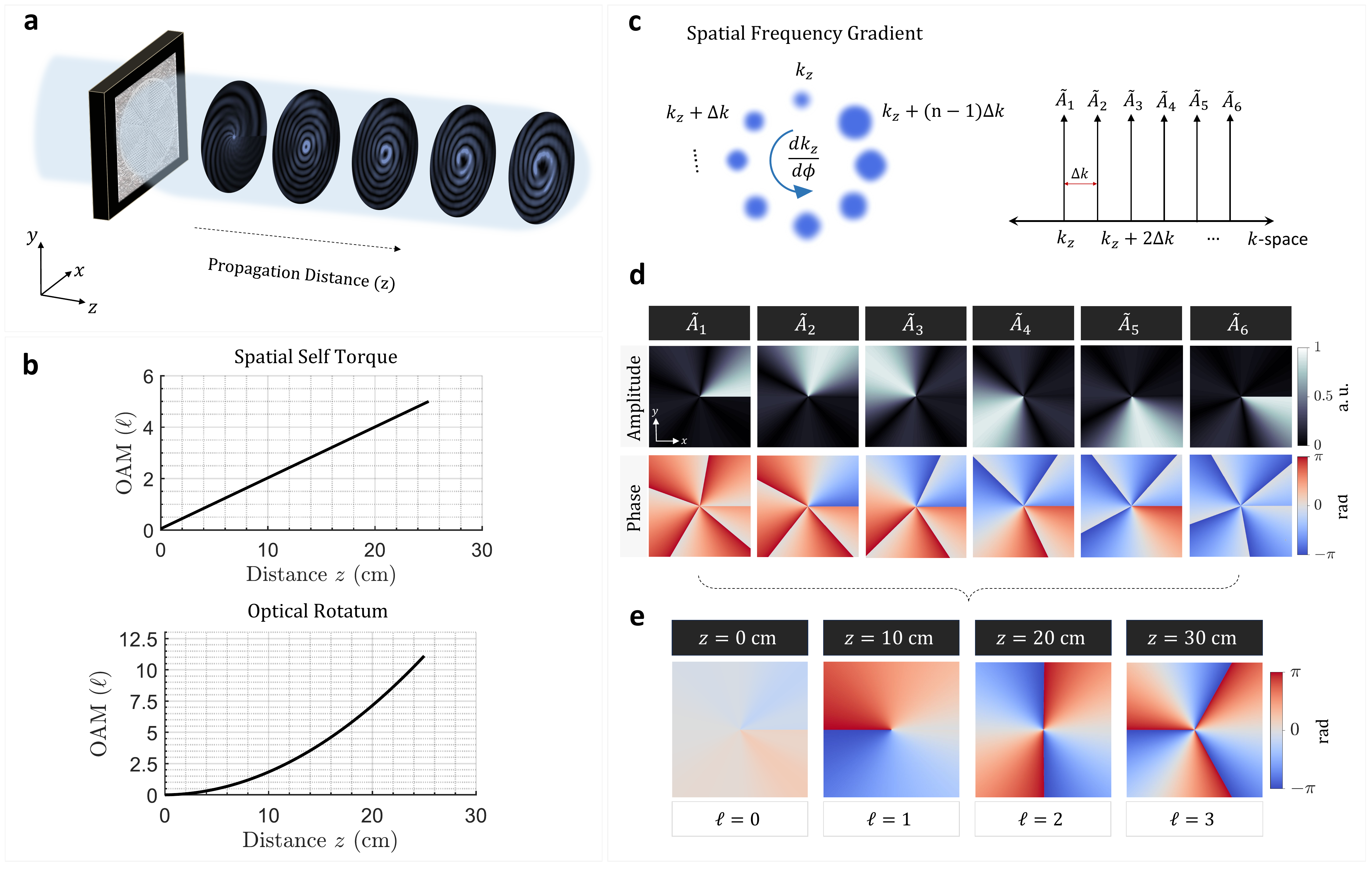}
    \caption{\textbf{Rotatum of light}. \textbf{a)} A phase mask converts a plane wave into a vortex beam whose OAM can grow (or decay) following a quadratic dependence along the direction of propagation. \textbf{b)} The OAM, signified by $\ell$, can follow any arbitrary $z$-dependent profile which can be linear, quadratic, or cubic. The linear and quadratic evolution of OAM give rise to the spatial self-torque (top) and rotatum of light (bottom), respectively. \textbf{c)} The mechanism relies on introducing an azimuthal gradient in the spatial frequency of the beam; \Rev{$dk_z/d\phi$} (left). This can be realized by creating a spatial frequency comb in the $k_z$ domain. Each comb tooth is weighted by a complex coefficient, $\tilde{A}_n$ (right). \textbf{d)} Amplitude and phase profiles of the coefficients $\tilde{A}_n$, designed in this case to produce a vortex with linearly evolving OAM. Each coefficient, $\tilde{A}_n$, is associated with a different spatial frequency, $k_{z,n}$. \textbf{e)} Upon propagation, different components of the beam (i.e., its $k$-vectors) weighted by $\tilde{A}_n$ will interfere forming an envelope with unity amplitude and $z$-dependent phase profile which adiabatically deforms its helical twist along the optical path via spatial beating.}
    \label{Fig1}
\end{figure}

We seek a phase mask which converts an incident plane wave into a vortex beam that changes its OAM in a parabolic manner along the optical path, as illustrated in Fig.~\ref{Fig1}(a). The OAM ($\ell$) shall evolve continuously following a linear, quadratic, or even cubic $z$-dependence. Here, we focus on linear and quadratic OAM evolution as depicted in Fig.~\ref{Fig1}(b). The former can be interpreted as a spatial self-torque of light whereas the latter is its rotatum. Tailoring the evolution of OAM to follow a parabolic $z$-dependence requires the wavefront to change its helical twist continuously as light propagates. In other words, any two points in the azimuthal direction should accumulate slightly different phase shifts upon propagation. To achieve this, we introduce an azimuthal gradient in the spatial frequency of the beam. Consequently, the propagation constant will also vary, point-by-point, in the azimuthal direction. As the beam propagates, it acquires a helical wavefront which continuously deforms its twist following any predetermined profile. Figure~\ref{Fig1}(c) illustrates this concept: A discrete set of monochromatic wave sources with different propagation constants are arranged in a ring formation. Each of these sources (or modes) has a different $k_{z,n}$ vector, with equal separation in $k_z$-space, and is weighted by different complex coefficients, $\tilde{A}_n$. Here, $k_z$ denotes the longitudinal component of the wavevector and $n$ is the index of each azimuthal mode. 

The coefficients $\tilde{A}_n$ depend on the target OAM profile. For example, let the desired vortex evolve following $e^{i\ell(z)\phi}$ dependence (where $\ell$ is now a function of $z$) over a finite distance of $L$. To obtain $\tilde{A}_n$, we solve the following Fourier integral:
\begin{equation}
    \tilde{A}_n = \frac{1}{L}\int_{0}^{L} e^{i\ell(z)\phi} e^{-i \frac{2\pi n}{L}z}dz.
\label{Eq1}
\end{equation}
Equation~(\ref{Eq1}) yields a discrete set of 2D phase and amplitude profiles whose superposition constructs the target function $e^{i\ell(z)\phi}$. This is specifically true if the $k$-vector separation associated with two consecutive modes, $\tilde{A}_n$ and $\tilde{A}_{n-1}$, is $k_{z,n}-k_{z,n-1}=2\pi/L$. 

The first six coefficients $\tilde{A}_n$ for a wavefront that evolves as $\sim e^{i10z\phi}$ are plotted in Fig. 1(d). When added together and propagated the resulting envelope will acquire a \Rev{quasi-uniform} phasefront at $z=0$, and helical phasefronts with $\ell=1$, and $\ell=2$, at $z=10$ cm and $z=20$ cm, respectively, as shown in Fig.~\ref{Fig1}(e). Hence, by substituting any target OAM profile in the Fourier integral of Eq.~(\ref{Eq1}), one can find the coefficients $\tilde{A}_n$ of each azimuthal mode. 
The next step is to determine the profile of our propagating modes (i.e, the spatial carriers) which provide the $\sim e^{ik_{z.n}z}$ dependence. Bessel beams are ideal candidates for this purpose given their non-diffracting and self-healing behavior~\cite{McGloin2005}. Traditionally constructed by axicons or plane waves along a cone, Bessel beams have a spatial frequency (propagation constant) which can be precisely tuned by changing their cone angle---a feature that perfectly fits our approach. Therefore, the profile of our proposed optical vortices is mathematically expressed as~\Rev{\cite{Dorrah2021NC}}:
\begin{equation}
    \psi(\rho,z,t) = \sum^{n=N}_{n=-N} \tilde{A}_n J_0(k_{{\rho,n}}\rho)e^{i k_{z,n} z}e^{-i \omega t}.
    \label{Eq2}
\end{equation}
\noindent Here, $J_0$ denotes the Bessel function of the first kind, $k_\rho$ and $k_z$ are the transverse and longitudinal wavenumbers, respectively, and the term $e^{-i\omega t}$ denotes the harmonic time dependence. The summation consists of 2$N$+1 modes. The mathematical formulation of Eq.~(\ref{Eq2}) is treated more rigorously in the Methods Section. By solving this equation at $z=0$, one can obtain the 2D profile of the field distribution which, upon propagation, will acquire the desired helical phase profile $e^{i\ell(z)\phi}$. 

\section{Results}

Substituting $z=0$ in Eq.~(\ref{Eq2}) provides the complex field profile that shall be implemented on the wavefront shaping tool of choice; for e.g., metasurfaces or spatial light modulators (SLMs). To generate our target vortex profiles, we used a standard holography setup composed of a phase-only reflective SLM and $4$-$f$ imaging system as depicted in Fig.~\ref{Fig2}(a) and described more fully in the Methods. In the following, we demonstrate the experimental generation of optical vortices in which the OAM varies linearly and quadratically along the optical path. We also discuss the underlying physical dynamics associated with such evolution.
 
\subsection{Linearly growing vortices:} We start with a scenario in which a vortex beam experiences a linear chirp in its OAM as it propagates. To achieve this, the vortex beam shall acquire a helical phasefront that is a linear function of $z$, following $\sim e^{i\ell(z)\phi}$ dependence. The chirp rate of $\ell$ as well as its sign can be designed at-will to create a vortex with linearly increasing (or decreasing) OAM. A linear monotonic variation in OAM as a function of $z$ is the spatial self-torque, as described in the Introduction. To be specific, consider a beam in which the  target helical phase is given by $\sim e^{i10\phi z}$. The coefficient $10$ is a normalization constant which defines the linear chirp rate of the azimuthal phase---i.e., slope of the linear OAM evolution---along the optical path. This particular choice of chirp rate allows the beam to \Rev{increase its OAM ($\ell$) by an integer value every $10$ cm}. It could have been set to other values as well. To realize the desired vortex beam, we evaluate $\tilde{A}_n$ by solving Eq.~(\ref{Eq1}) then we obtain the initial field distribution from Eq.~(\ref{Eq2}). By doing so, we find that the coefficients $\tilde{A}_n$ are given by the closed form expression
\begin{equation}
    \tilde{A}_n = - \frac{e^{2\phi i - (\frac{4\pi n}{5})i -i}}{5\phi-2\pi n}.
\label{Eq3}
\end{equation}
\noindent The six most significant terms of $\tilde{A}_n$ are the same as the ones depicted in Fig.~\ref{Fig1}(d), exhibiting six amplitude masks with different azimuthal orientations. Choosing a different chirp rate for OAM would change the number of these azimuthal masks, producing finer or coarser sectors, which translates to modifying the azimuthal gradient in the $k$-vector. This is the case since each $\tilde{A}_n$ will be multiplied by a Bessel beam with different $k_z$. By substituting $\tilde{A}_n$ in Eq.~(\ref{Eq1}) and evaluating the expression at $z=0$, we obtain the 2D field distribution that shall be \Rev{encoded} on the SLM. A plane wave incident on the SLM would thus be transformed to a vortex beam whose topological charge follows the linear dependence $\ell=10 z$. 

\begin{figure}[h!]
    \centering
    \includegraphics[width=0.925\textwidth]{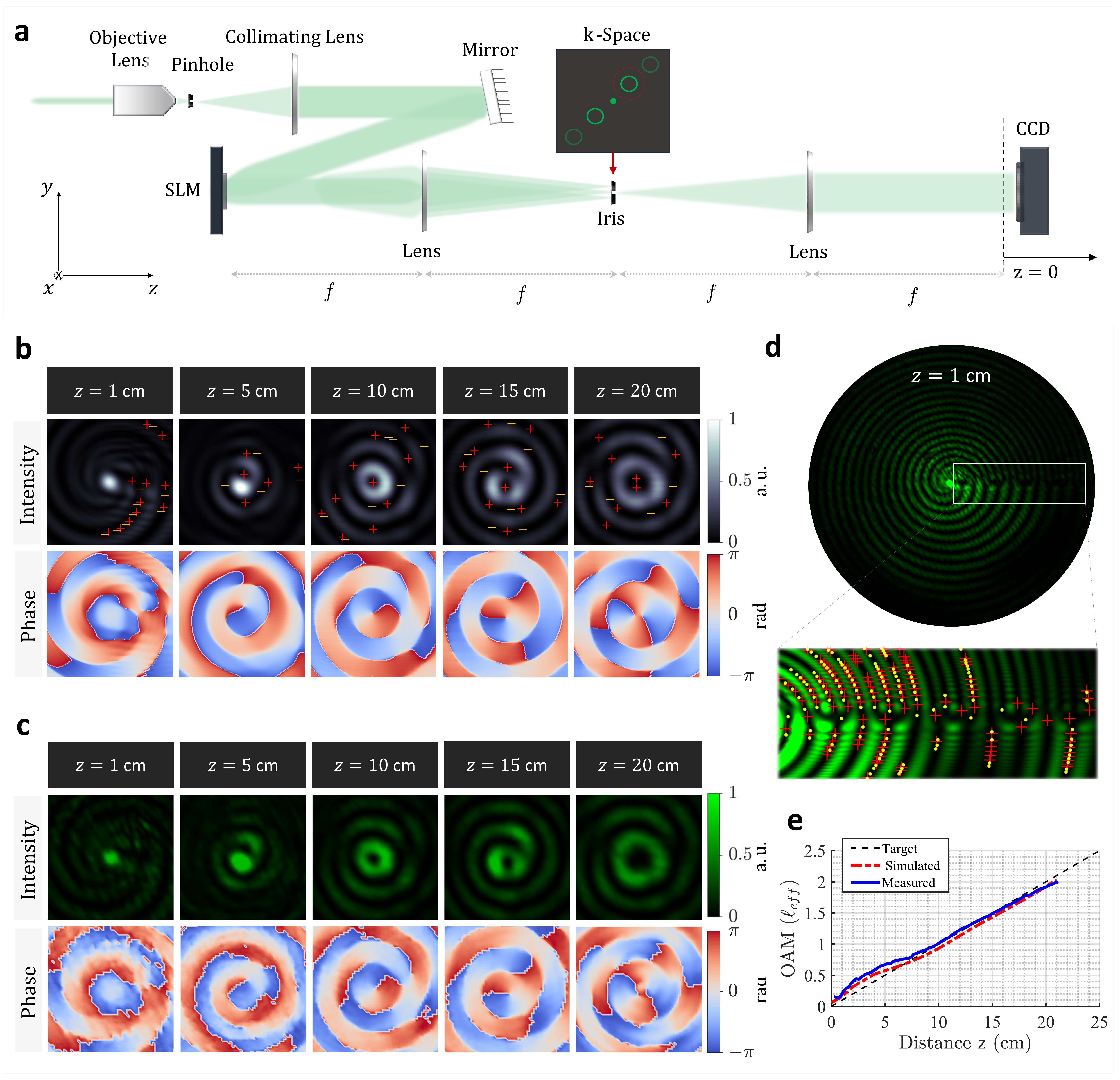}
    \caption{\textbf{Generation of a vortex beam with spatial self-torque}. \textbf{a)} Experimental setup: an expanded and collimated laser beam is incident on a reflective phase-only SLM with the desired hologram. The reflected beam is then filtered and imaged onto a CCD using a $4$-$f$ lens system. The CCD is mounted on a translation stage to capture the transverse profile of the beam at different $z$-planes. \textbf{b)} Simulated 2D profiles of the intensity and phase of a vortex beam in which the OAM increases linearly along the optical path. \Rev{The red and yellow markers denote phase singularities with positive and negative handedness, respectively}. \textbf{c)} Measured 2D intensity and reconstructed phase profiles of the beam in (b). \textbf{d)} The initial beam profile (at $z=1$ cm) exhibits a horizontal line of darkness and bifurcation. The inset depicts the underlying chain of phase singularities associated with this \Rev{dislocation} line. \textbf{e)} Measured and simulated evolution of OAM (effective topological charge) as a function of the propagation distance. } 
    \label{Fig2}
\end{figure}

Figure~\ref{Fig2}(b) depicts the simulated 2D intensity and phase profiles of the resulting beam at different propagation distances. The profiles were obtained using the Kirchhoff–Fresnel diffraction integral. Notice how the local wavefront slowly evolves from a flat to a helical profile as  the beam propagates. The \Rev{positive and negative} phase singularities in the vicinity of the beam's center are denoted by the red and yellow markers. They signify local vortices of opposite handedness. The spatial movement of these singularities play a role in modifying the topological charge from $\ell=0$ at $z=1$ cm to $\ell=1$ at $z=10$ cm and $\ell=2$ at $z=20$ cm. The measured intensity and phase profiles are shown in Fig.~\ref{Fig2}(c) and are in very good agreement with the simulated ones in Fig.~\ref{Fig2}(b). The phase has been reconstructed from the intensity measurements using the single-beam multiple-intensity reconstruction (SBMIR) technique~\cite{Almoro:06} described in Supplementary Section 4. \Rev{The  evolution of the beam's phase and intensity profiles is captured in Supplementary Video 1.}

A closer look at the initial beam profile (at $z=1$ cm), depicted in Fig~\ref{Fig2}(d), reveals two key features of this vortex: a) a region of bifurcation or \Rev{dislocation~\cite{doi:10.1098/rspa.1974.0012}} due to the mismatch in the spatial frequency along the azimuthal direction. \Rev{This discontinuity arises precisely where the fast and slow spatial oscillations of the Bessel beam merge as depicted in Fig.~\ref{Fig1}(c)}, and b) a chain of phase singularities (i.e, local vortices) of alternating polarity. This chain of vortices is a universal feature of OAM beams with fractional topological charge~\cite{doi:10.1098/rspa.1974.0012,MVBerry_2004,ZhangZeng}. 

To confirm the linear growth of OAM, we calculate the effective topological charge ($\ell_{\text{eff}}$) of the beam~\cite{{Allen1999,Litvin:11,Schulze_2013}}. In essence, $\ell_{\text{eff}}$ is proportional to the ratio between the total OAM and energy of a given field distribution. \Rev{It is expressed as
$\ell_{\text{eff}} = {\omega}{\mathbf{L}_z}/{\mathbf{W}}$ and provides a quantitative measure of the OAM per photon~\cite{Litvin:11,Schulze_2013}}. The derivation for calculating effective topological charge can be found in the second subsection of the Methods. Figure~\ref{Fig2}(e) depicts the simulated and experimentally evaluated $\ell_{\text{eff}}$ as a function of propagation distance in comparison with the target linear profile. Here, $\ell_{\text{eff}}$ has been evaluated locally by encircling a finite region around the beam's center. From this result, we infer that the OAM is locally chirped along the optical path following the desired linear profile. \Rev{This effect is different from previous demonstrations~\cite{PhysRevA.98.043846,Dorrah2021NC} in which $\ell_{\text{eff}}$ was allowed to acquire integer values or weighted superpositions thereof}. Notably, linear chirp in $\ell_{\text{eff}}$ occurs only locally at the center of the beam. The global OAM is always conserved aided by the movement of the phase singularities (local vortices) across the beam~\cite{PhysRevA.98.043846,Dorrah2021NC}. This mechanism is general and can allow other types of OAM evolution which can be non-monotonic or even parabolic as we will show. In the next example, we demonstrate another set of vortices whose charge follows a non-monotonic evolution (i.e., growth and decay) and we point to an underlying topological phase factor that accompanies such transition.

\subsection{Non-monotonic linear OAM evolution}

\begin{figure}[h!]
    \centering
    \includegraphics[width=0.975\textwidth]{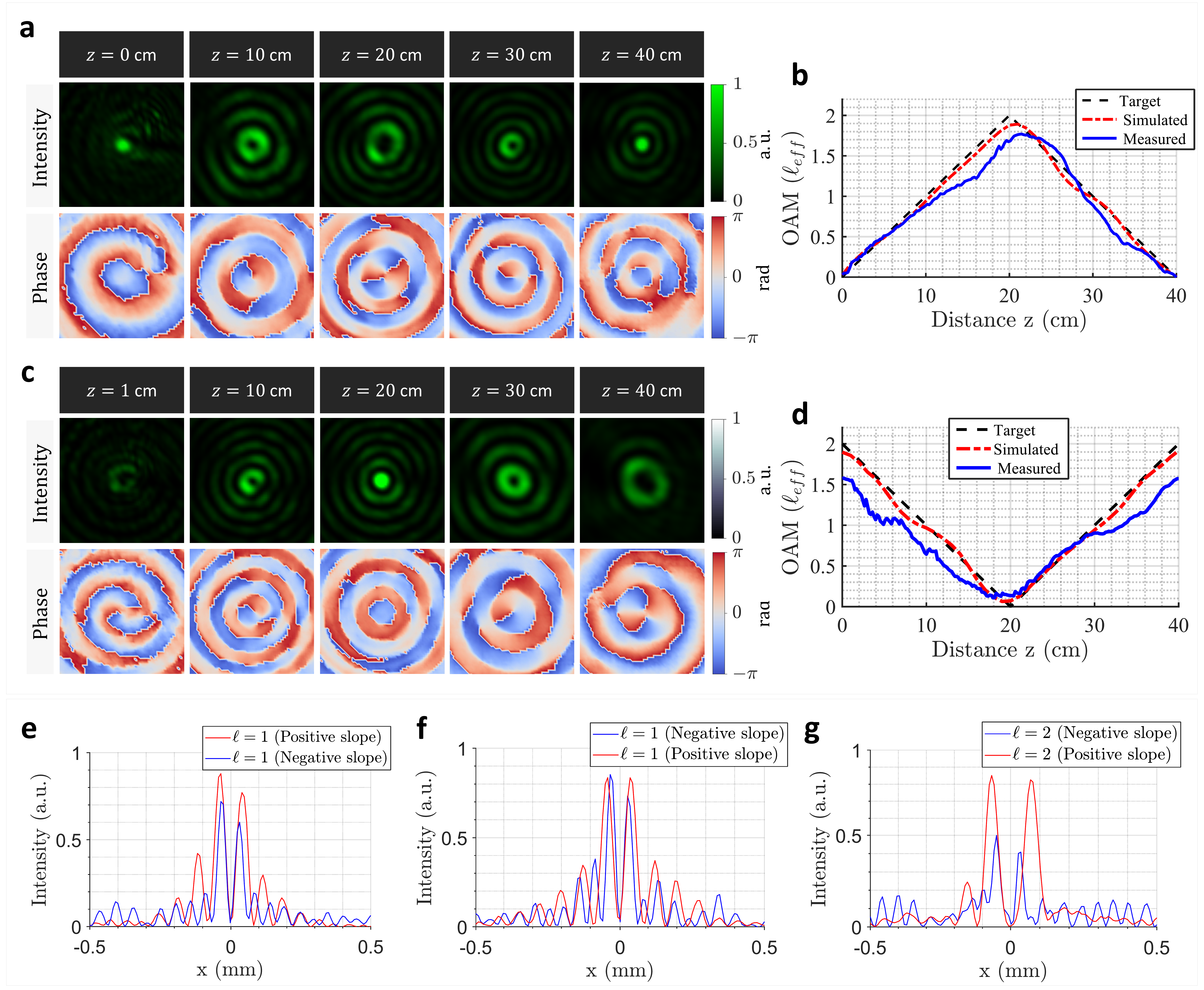}
    \caption{\textbf{Optical vortices with non-monotonically varying charge}. \textbf{a)}. Measured transverse profiles (intensity and phase) of a vortex beam whose OAM locally grows then decays along the optical path. The beam acquires a helical phase, increasing its OAM adiabatically from $\ell=0$ to $\ell=2$, in a linear manner over a range of $20$ cm. Afterwards, the local OAM decreases continuously from $\ell=2$ to $\ell=0$ as the beam propagates for longer distance. \textbf{b)} Measured and simulated spatial evolution of OAM (effective charge, $\ell_{\text{eff}}$) in comparison with the target design for the optical vortex in (a). \textbf{c)} Measured 2D intensity and phase profiles of an OAM beam whose vorticiy decays then grows with propagation. The beam starts with a local charge of $\ell=2$ then slowly unwinds its helicity to $\ell=0$ before it acquires the same helical phase ($\ell=2$) again, albeit with different size. \textbf{d)}. The measured and simulated evolution of the corresponding charge ($\ell_{\text{eff}}$) as a function of $z$. Vertical 1D cuts of the transverse profiles in (a) and (c) are plotted in \textbf{(e-g)}, respectively. These cuts suggest that \Rev{the beam's size is perturbed (i.e., it experiences a $k$-shift) even when the topological charge is the same. The change in size depends on whether $\ell_{\text{eff}}$ increases or decreases with propagation. }}
    \label{Fig3}
\end{figure}

While it seems counter-intuitive, a vortex beam which reverses its torque can be realized using our approach. This can be done by superimposing two sets of vortices; where the first experiences a linear growth in OAM (as shown before) whereas the second experiences a decay. Vortices with decaying OAM can be realized by setting the target phase profile to $e^{i(\ell_{\text{o}}-\ell(z))\phi}$, where $\ell_{\text{o}}$ is the initial charge of the beam and $-\ell(z)$ signifies the negative slope of charge evolution. To demonstrate this further, we present two scenarios in which a vortex beam experiences a linear growth then decay in its OAM and vice versa. Figure~\ref{Fig3}(a) depicts the measured 2D intensity and phase profiles for the first case: a pencil-like beam with a localized spot and uniform phasefront at the center slowly acquires a helical phase, evolving to a vortex with varying strength and diameter, before it rewinds its helicity and transforms back to a pencil beam. Notably, while the beam carries the same charge, $\ell=1$, at $z=10$ cm and $z=30$ cm, the beam's diameter is slightly perturbed at these positions. This effect is predicted from the simulated profiles based on Kirchhoff's propagation (see Supplementary Fig. 1) and will be explained shortly. The corresponding evolution of OAM (local charge) is shown in Fig.~\ref{Fig3}(b) confirming the linear dependence on $z$. \Rev{The dynamical evolution of the same beam is further captured in Supplementary Video 2.}

To contrast this picture, we generated another vortex whose local OAM decays then grows linearly with propagation. Figure~\ref{Fig3}(c) and \Rev{Supplementary Video 3} show the measured transverse intensity and phase profiles for this case. Here, a vortex beam with $\ell=2$ continuously unwinds its helical wavefront until it becomes locally uniform at $z=20$ cm before it acquires a helical phase again, increasing its charge to $\ell=2$. Similar to the case of monotonic OAM growth and decay, here the initial and final beam sizes (at $z=1$ cm and $z=40$ cm) are not the same despite carrying equal charge of $\ell=2$. To reconcile this behavior, recall that in order to modify the charge $\ell$, the phasefront should be continuously deformed with propagation. This occurs if the wavefront acquires an additional ($\phi$-dependent) phase factor with respect to a reference vortex with constant charge. In this case, the phase dependence of the ensemble can be expressed as
\begin{equation}
\Rev{e^{iK_z z} \sim e^{i(k_{z,0})z+i\ell_{\text{eff}}(z)\phi}},
\end{equation}
where $K$ is the effective longitudinal wavevector of the envelope which now has two contributing terms: i) a propagation phase term, $e^{ik_{z,0}z}$, as expected from a normal Bessel beam, and ii) an additional phase factor \Rev{$e^{i\ell_{\text{eff}}(z)\phi}$}. Hence, points along the azimuthal direction ($\phi$) of the wavefront acquire a $z$-dependent phase \Rev{$\Phi(z) = \ell_{\text{eff}} (z)\phi$}. The slope of this phase with respect to $z$ is reminiscent of an effective momentum, \Rev{which we denote as $k_{\text{B}}$}. Therefore, the propagating beam experiences a $k$-shift of \Rev{$k_{\text{B}} = {\partial{\Phi(z)}/\partial z}$} such that its effective longitudinal wavevector becomes \Rev{$K_z = k_{z,0} + k_{\text{B}}$}. The temporal frequency of the beam is unchanged and $K_z^2 + k_{\rho}^2 = (\omega/c)^2$; thus, from momentum conservation, a shift in $K_z$ mandates a subsequent shift in $k_{\rho}$. The latter translates to a change in the transverse size of the beam. Notably, the beam's size is perturbed depending on the sign and magnitude of $\ell_{\text{eff}}(z)$. This is consistent with the measured profiles of Fig.~\ref{Fig3}(a,c). To better visualize this effect, we plotted the 1D cuts of these transverse profile as shown in Fig.~\ref{Fig3}(e-g). These plots suggest that the linear growth and decay of OAM is associated with red and blue shifts in the spatial frequency of the beam, respectively. This dependence stems from our phase convention ($e^{ik_{z,0}z}$) and would be reversed if the $e^{-jk_{z,0}z}$ convention is adopted instead. A vortex beam with constant charge does not experience such a perturbation in its $k$-vector (see Supplementary Fig. S1). An analogue of this additional phase factor (and associated $k$-shift) has been observed in the case of vector beams with $z$-dependent polarization evolution~\cite{Dorrah2021NP}. The latter has been reconciled as a Berry phase factor which is acquired as the beam adiabatically modifies its spin angular momentum with propagation. Similarly, here, as the vortex beam undergoes an adiabatic evolution in its parameter space ($\ell$), to deform its topology, \Rev{it acquires an additional phase factor which can be reconciled as a Gouy~\cite{Martelli:10} or Berry phase~\cite{Berry_Phase,Berry2010,Cohen2019}}. In the following, we expand the scope of our method and demonstrate vortex beams with quadratic evolution of OAM.

\subsection{Quadratic evolution of OAM}

The slope and curvature of the OAM evolution can be designed on demand using our approach. By allowing the topological charge to experience a quadratic chirp along the optical path, we create a vortex beam with optical rotatum. To demonstrate this, we consider a vortex beam with a helical phasefront that follows $\sim e^{i100z^2\phi}$ dependence. This allows the vortex beam to reach $\ell=1$, $2$, $3$, and $4$ at $z=10$,  $14$, $17$, and $20$ cm, respectively. Setting this as the target function in Eq.~(\ref{Eq1}) with $L=50$ cm yields the following closed form expression for $\tilde{A}_n$:
\begin{equation}
    A_n = i\frac{\sqrt{\pi}e^{-\frac{i n^2 \pi^2}{25\phi}}erf\left ( \frac{2\pi n}{\sqrt{100i \phi}} \right )}{\sqrt{100i\phi}}+i\frac{\sqrt{\pi}erf\left ( \frac{40 \phi - 2 \pi n}{\sqrt{100i\phi}} \right )e^{-\frac{ i n^2\pi^2}{25 \phi}}}{\sqrt{100i\phi}}.
\end{equation}

\noindent These are the complex amplitude terms of the Bessel beams in Eq.~(\ref{Eq2}). Solving the latter at $z=0$ provides the target profile to be generated by the SLM.

\begin{figure}[h!]
    \centering
    \includegraphics[width=0.9\textwidth]{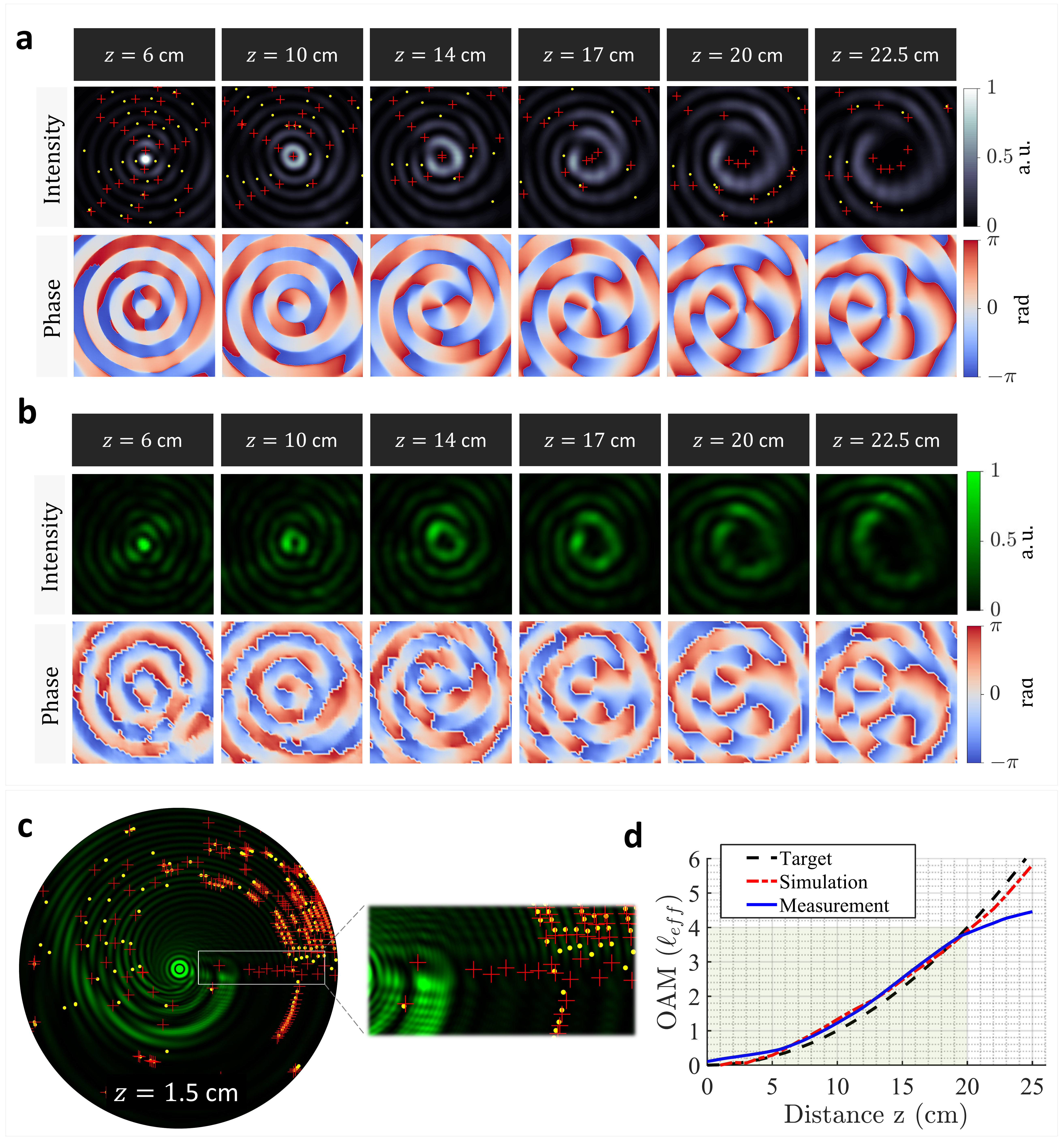}
    \caption{\textbf{Optical vortices with quadratic evolution of topological charge: optical rotatum.} \textbf{a)}. Simulated transverse profiles (intensity and phase) of a vortex beam whose OAM locally grows in a quadratic manner along the optical path. The beam acquires a helical phase, increasing its OAM continuously from $\ell=0$ to $\ell=5$ over a range of $22.5$ cm. The \Rev{red and yellow markers denote phase singularities of opposite handedness (positive and negative helicity, respectively)}. \textbf{b)} Measured 2D intensity and phase profiles at the $z$-planes in (a). \textbf{c)} The intensity profile of the vortex at $z=1.5$ cm. The inset depicts a close up exhibiting a line of phase singularities feeding the beam's center. \textbf{d)} Comparison between the target, simulated, and measured local charge ($\ell_{\text{eff}}$) showing its quadratic dependence on $z$.}
    \label{Fig4}
\end{figure}

Figure~\ref{Fig4}(a) shows the simulated 2D intensity and phase profiles of the resulting beam at different propagation distances. In this case, the wavefront evolves from a locally flat ($\ell=0$) to a helical ($\ell=5$) profile, in a continuous manner, as the beam propagates. The red and yellow markers denote the positive and negative phase singularities. The precise movement of these singularities underpins the evolution of the topological charge at different $z$-planes. For instance, at each $z$-plane, we notice that an additional singularity is accumulated inside the central ring of the beam. The singularities approach the beam's center via the dark fringes in its vicinity. The measured intensity and phase profiles are shown in Fig.~\ref{Fig4}(b) and \Rev{Supplementary Video 4} and are in very good agreement with the calculated ones. The initial beam profile (at $z=1$ cm) is depicted in Fig~\ref{Fig4}(c). Similar to the case of vortex beam with linear evolution of OAM, here we observe a few key features: a) a region of bifurcation at the interface between the fast and slow spatial oscillations of the Bessel beam. This \Rev{dislocation} arises due to the local mismatch in the spatial frequency along the azimuthal direction \Rev{(see for e.g., Fig.~\ref{Fig1}(c))}. b) A chain of phase singularities (i.e, local vortices) of alternating signs which approach the beam's center via a dark intensity line. 

\noindent This is a well known signature of beams with fractional topological charge as previously discussed. Additionally, a distribution of phase singularities is observed at the outer peripherals of the beam. Its contour follows a growing logarithmic spiral and is reminiscent of the shape of a Nautilus seashell (See Extended Data, Fig.2). 
Interestingly, the growth factor of this spiral is very close to Fibonnaci's golden ratio \Rev{which is also close to our target quadratic growth for $\ell_{\text{eff}}$ where $e^{i\ell_{\text{eff}}\phi}\sim e^{i\phi 100z^2}$}. 

To confirm the quadratic of OAM, we calculated and measured the local effective topological charge ($\ell_{\text{eff}}$) of the beam as a function of propagation distance. A comparison between the target and realized profiles is depicted in Fig.~\ref{Fig4}(d). From this result, we infer that the OAM is locally chirped along the optical path following the desired parabolic profile. Notably, this behavior occurs only locally at the center of the beam. The global OAM is always conserved owing to the engineered movement of the phase singularities across the beam~\cite{PhysRevA.98.043846,Dorrah2021NC}. Note that, in analogy with the linearly evolving OAM, the quadratic evolution here is also associated with an accumulated Berry phase factor which perturbs the spatial frequency of the beam. In this case, the $k_z$-vector of the ensemble experiences a linear chirp along the optical path. We examine this effect further in Extended Data Fig. 1. Furthermore, we simulated another scenario of optical rotatum with a larger number of Bessel beams which better approximate the parabolic evolution of OAM (Supplementary Fig.2).  Lastly, the temporal analogue of optical rotatum can be realized using Bessel beams of different wavelengths such that their beating gives rise to a time varying OAM. This will be the subject of future work.

\section{Discussion and Outlook}
We introduced a new degree-of-freedom of light, optical rotatum, in which an optical vortex experiences a chirp in its OAM as it propagates. We showed that such an adiabatic deformation of the topological charge is associated with a Berry phase factor that perturbs the spatial frequency of the beam. Our approach is fully-analytical and can be applied to other regions of the electromagnetic spectrum in addition to ultrasound and electron beams. Structured waves of this kind may inspire new directions in science and technology. It advances the field of singular optics by enabling topologically complex states of light which in turn can lead to many new phenomena in quantum and classical optics~\cite{Fickler640,Hancock:19}. More specifically, spatially chirped vortices can be used in depth sensing, metrology, and free space communications. Furthermore, as an unexplored property of light, rotatum may reveal a new class of optical forces which can be exploited in light-matter interaction, micromanipulation, and spintronics both in the near and far-field regimes. 

It is noteworthy that given our choice of Bessel functions as the OAM modes, our vortex beams are characterized by a non-diffracting and self-healing behavior~\cite{McGloin2005} which is desirable in many applications. Moreover, the dimensions of our beams as well as their propagation range can be readily modified by changing the aperture size of the devices and the cone angles of the OAM modes following the same design considerations of axicons~\cite{McGloin2005}. While we primarily focused on scalar vortex beams, our approach can be extended to vector beams with spatially-varying polarization states, enabling rich spin-orbit interactions in free space. This will be the subject of future work. Lastly, the multidisciplinary nature of angular momentum and singularity engineering across different fields may inspire related research efforts in the areas of microfluidics, acoustics, and pattern formation, to name a few. Therefore, we thus envision this work to enrich the science and applications of structured light and beyond. 

\pagebreak
\section*{Methods}

\subsection{Engineering the Vortex Beams:}
The spatially evolving optical vortices discussed in this work were constructed from the superposition of co-propagating Bessel vortex beams given by
\begin{equation}
    \psi(\rho,z,t) = \sum^{n=N}_{n=-N} \tilde{A}_n J_0(k_{{\rho,n}}\rho)e^{i k_{z,n} z}e^{-i \omega t},
\end{equation}
\noindent where $k_{\rho,n}$ and $k_{z,n}$ denote the transverse and longitudinal components of the wavevectors, respectively. For the case of optical vortices with linear evolution of OAM, we set $N=8$ which yields 17 Bessel beams in the superposition. The $17$ Bessel beams are equally spaced in $k_z$-space by a separation of $2\pi/L$. The central $k_{z,0}$ was set to $0.999991$ $k_0$ and the resulting beam extended for a range $L= 50$ cm. These parameters dictate the degree of paraxiality of the generated beam. For the case of vortices with quadratic evolution of OAM, we reduced $k_{z,0}$ to $0.0.99998$ $k_0$ and in turn increased $N$ to $18$. This allowed us to include $36$ Bessel beams in the superposition, providing a more accurate reconstruction of the target evolution. To ensure effective generation of the target vortex profile over the propagation range $L$, with minimal diffraction, the aperture radius of the wavefront shaping device should satisfy the following criterion~\cite{Dorrah2021NC}
\begin{equation}
    R_{\text{aperture}}= L\sqrt{\bigg[\frac{k_0}{ ({k}_{z,n})_\text{max}}\bigg]^2-1}.
    \label{EqS2}
\end{equation}

\subsection{Evaluating the Effective Topological Charge:}  The topological charge, $\ell$, is typically a quantized value. Nevertheless, given a local field distribution, one can calculate an effective topological charge ($\ell_{\text{eff}}$) by evaluating the OAM and energy densities of that field. This quantity can be integer or fractional. We start by deriving an expression for the OAM density following Refs.~\cite{Litvin:11,Schulze_2013}. In this case, the time-averaged Poynting vector is expressed as
\begin{equation}
\mathbf P = c^2 \epsilon_{0} \frac{1}{2}\Re\{\mathbf{E}\times \mathbf{B}^*\}. 
\label{EqS7}
\end{equation}
\noindent The magnetic flux density, $\mathbf B$, can be written in terms of the electric field through Maxwell-Faraday equation ($\nabla \times \mathbf{E} = - i \omega \mathbf{B}$). The Poynting vector then becomes
\begin{equation}
\mathbf P = \frac{\epsilon_{0}c^2}{2 \omega} \Re\{i \mathbf{E}\times (\nabla \times \mathbf{E})^*\}. 
\label{EqS7}
\end{equation}
Here, $\epsilon_{0}$ is the free space permittivity ($8.854 \times 10^{-12}$ F/m), $\omega$ is the angular frequency, and $c$ is the speed of light in vacuum. The OAM density is then evaluated from
\begin{equation}
\mathbf{j} = \frac{1}{c^2} (\mathbf{r}\times \mathbf P), 
\label{EqS7}
\end{equation}
\noindent where $\mathbf r$ is the position vector ($\mathbf r = \text{x} \hat{x} + \text{y} \hat{y} + \text{z} \hat{z}$). The longitudinal component of OAM density ($\mathbf{j}_z$) is the quantity that is relevant to our purposes. Integrating $\mathbf{j}_z$ over a given transverse cross section of the beam yields the OAM associated with that area, denoted as $\mathbf{L}_z$ such that
\begin{equation}
\mathbf{L}_z = \int\int \mathbf{j}_z  rdr d\phi, 
\label{EqOAM}
\end{equation}
\noindent where it is understood that $\mathbf{L}_z$ is evaluated per unit length. To evaluate the effective charge, however, we need to normalize $\mathbf{L}_z$ by the total energy of the beam. In this case, the energy density is readily obtained from the Poynting vector; $w = c \epsilon_{0} \frac{1}{2}\Re\{E\times B^*\}$. The total energy per unit length, $\mathbf{W}$, is then obtained by integrating $w$ over the transverse profile of the beam such that
\begin{equation}
    \mathbf{W} = c\epsilon_0\frac{1}{2}\int\int \Re\{ E\times B^*\} rdrd\phi.
\label{EqS7}
\end{equation}
Normalizing $\mathbf{L}_z$ by $\mathbf{W}$ yields a quantity that is proportional to the mean OAM per photon, scaled by $1/(\hbar\omega)$; see for e.g., Eqs. (2.8) and (2.18) in Allen et al.~\cite{Allen1999}. Therefore, in the paraxial regime, the effective topological charge ($\ell_{\text{eff}}$) can be calculated from the ratio
\begin{equation}
    \frac{\mathbf{L}_z}{\mathbf{W}}=\frac{\ell_{\text{eff}}}{\omega}
\end{equation}

\subsection{Experimental Setup:} To generate the desired vortex beams, we used a reflective phase-only SLM (Santec SLM-200) with $1920 \times 1200$ pixel resolution and $8$ $\mu$m pixel pitch. We started by converting the 2D complex amplitude profile of Eq.~(\ref{Eq2}) to a phase-only computer generated hologram, to be compatible with our SLM, following the method outlined in Ref.~\cite{Arrizon:07}. Our measurements were obtained using a $532$ nm laser source (Novanta Photonics, Ventus Solid State CW laser) with the standard 4-$f$ holography setup depicted in Fig.~\ref{Fig2}(a). The laser beam was first expanded and collimated (using a 40X objective lens, a $100$ $\mu$m pinhole, and a $50$ cm lens) onto the reflective SLM screen. The desired complex amplitude spectrum was generated at the Fourier plane ($k$-space) of the SLM using a lens. The spectrum was then filtered from the zeroth and higher diffraction orders with an iris then transformed back to real space using a second lens. The latter imaged the generated vortex onto a charge-coupled device (CCD) camera (Thorlabs DCU224C, $1280\times1024$ resolution) which was mounted on a translation stage (Thorlabs LTS150) to capture the beam's evolution with steps of $0.25$ mm along its optical path. From these $z$-dependent intensity measurements, the phase profile was also retrieved following the single-beam multiple-intensity reconstruction (SBMIR) technique~\cite{Almoro:06}. To limit the noise effects during data acquisition, we used an adapted version of the flat fielding procedure described in the European Machine Vision Association's Standard 1288~\cite{EMVA1288}. The following adapted flat fielding procedure was previously performed in other work\cite{Li:23}. Profiles of the sensor's dark image and pixel-wise responsivity were captured and applied to each measurement taken. The same $532$ nm laser source was used as an input for sensor characterization. The power of the laser was driven from 0\% to 100\% in 5\% steps, where 100\% is the illumination level required to saturate the sensor at the exposure times used to capture each data set. The 0\% illumination image was taken as the dark current response of the sensor. To limit shot noise effects, between 50--70 frames were averaged per frame. Each pixel on the sensor then had its responsivity curve fit to the irradiance witnessed by the independent reference photodetector. The responsivity curve was inversely applied to the images before effective topological charge was calculated.

\subsection{Acknowledgements}
The authors wish to acknowledge S.W.D. Lim and J. Oh both of Harvard University for the useful discussion. A.H.D. acknowledges financial support from the Natural Sciences and Engineering Research Council of Canada (NSERC) under award no. PDF-533013-2019 and from the Optica Foundation Challenge program. F.C. acknowledges financial support from the Office of Naval Research (ONR), under the MURI program, grant no. N00014-20-1-2450, and from the Air Force Office of Scientific Research (AFOSR), under grant no. FA9550-22-1-0243.

\subsection{Author Contributions}
A.H.D. conceived the theory, developed the simulation, designed and built the experiment, and analyzed the data. A.P. created the computer generated holograms, performed the measurements, analyzed and processed the data. L.L. contributed to the data acquisition and analysis. F.C. supervised the project. All authors contributed to writing the manuscript.

\subsection{Data Availability}
All key data that support the findings of this study are included in the main article and its supplementary information. Additional data sets and raw measurements are available from the corresponding author upon reasonable request.

\subsection{Code Availability}
The codes and simulation files that support the figures and data analysis in this paper are available from the corresponding author upon reasonable request.


\subsection{Competing Interests} The author declare no competing interests.


\pagebreak
\bibliography{Bibliography}


\pagebreak

 \begin{figure}[h!]
    \centering

    \includegraphics[width=0.975\textwidth]{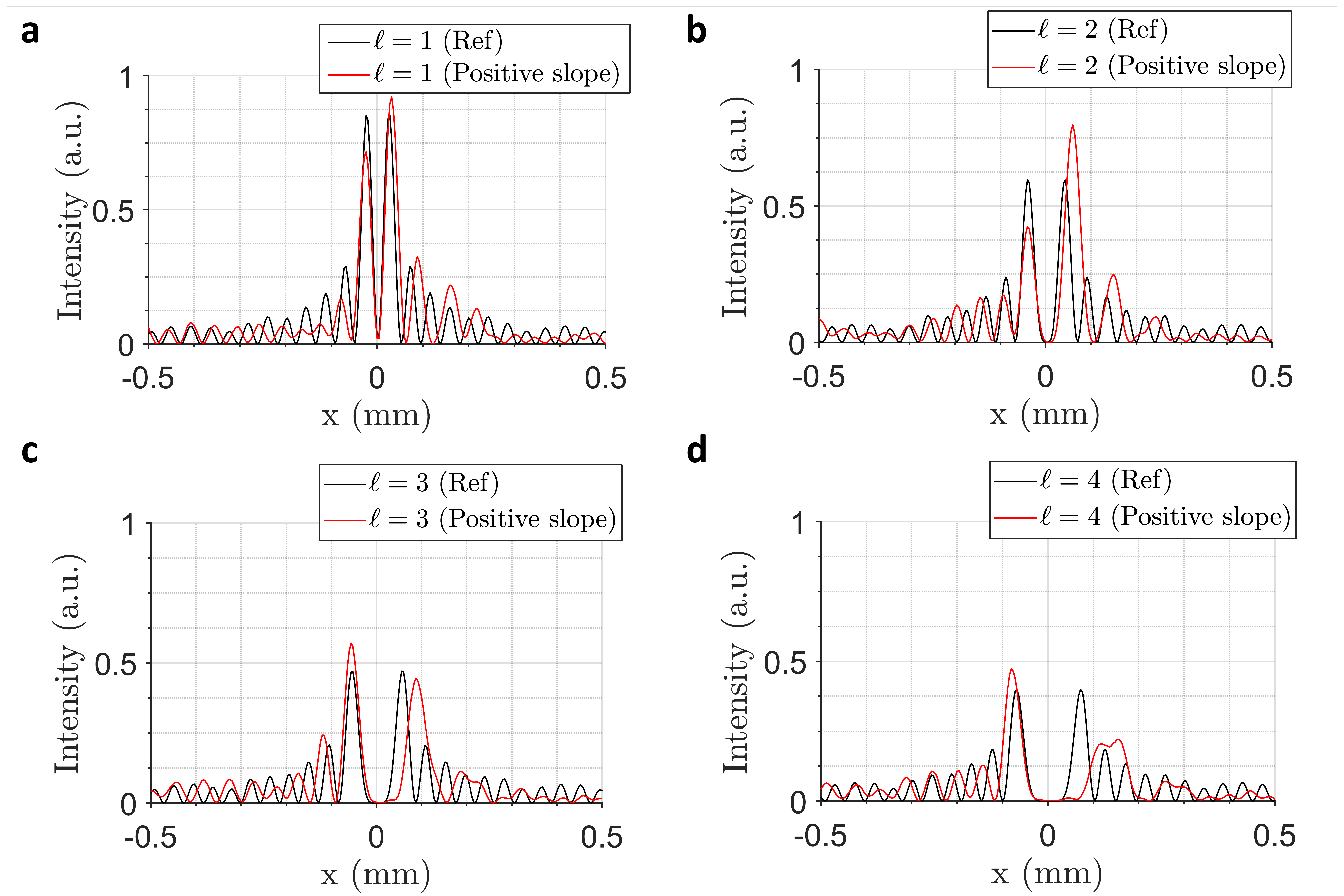}
    \caption*{\textbf{Extended Data, Fig. 1: Rotatum and spatial frequency shifts.} Simulated 1D transverse cuts of the intensity profile for the vortex beam of Fig. 4 in comparison to a vortex beam with constant OAM, $\ell$. Here, the spatially varying vortex changes its OAM in a quadratic manner along the optical path. Its profile is compared with a reference vortex of fixed $\ell$ at four different $z$-planes: $z=10$ cm (\textbf{a}), $14$ cm (\textbf{b}), $17$ cm (\textbf{c}), and $20$ cm (\textbf{d}). It is observed that the spatially evolving vortex experiences a red shift in its transverse spatial frequency ($k_{\rho}$) which can be inferred from the slight perturbation (stretching) in the lateral dimensions of the beam compared to a reference vortex. The underlying mechanism of this $k$-shift is an accumulated propagation-dependent Berry phase.}
    \label{Fig5}
\end{figure}

\pagebreak
 \begin{figure}[h!]
    \centering

    \includegraphics[width=0.975\textwidth]{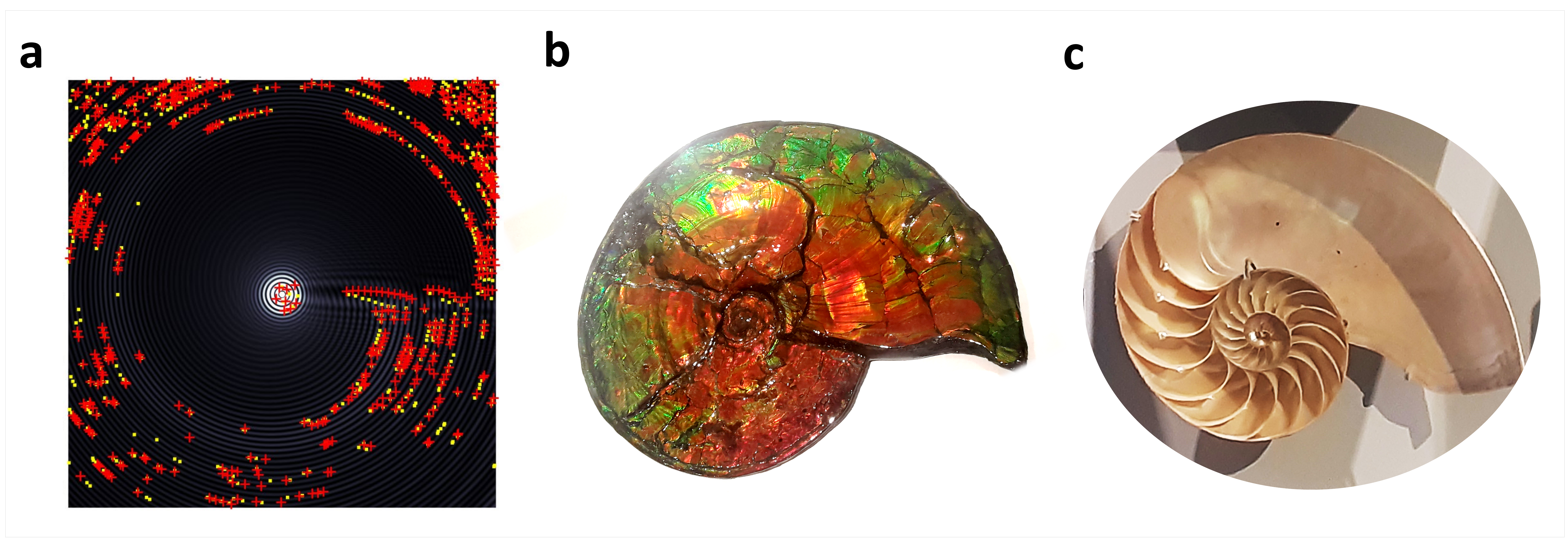}
    \caption*{\textbf{Extended Data, Fig. 2: Rotatum and logarithmic spirals.} \textbf{a)} Simulated 2D intensity profile of a vortex beam with optical rotatum at a propagation distance of $z=1.5$ cm. Here, the vortex is composed of 91 Bessel vortex beams ($N=45$). The red and yellow markers denote the positive and negative phase singularities. The contour of these singularities follows a logarithmic spiral pattern that resembles many phenomena in nature such as pattern formation in crystals and seashells. \textbf{b)} An image of Aragonite which is a carbonate mineral and one of the three most common naturally occurring crystal forms of calcium carbonate. Aragonite is formed by biological and physical processes, including precipitation from marine and freshwater environments. \textbf{c)} The chambered nautilus, also called the pearly nautilus, is the best-known species of nautilus. The shell, when cut away, reveals a lining of lustrous nacre and displays a nearly perfect equiangular spiral.}
    \label{Fig6}
\end{figure}

\end{document}